# An analysis of B meson decays to two light pseudoscalar nonets with $|\Delta S| = 1$ in QCD factorization approach

Janardan P. Singh[1] and N. G. Deshpande[2]
[1]Physics Department, Faculty of Science, The M. S. University of Baroda, Vadodara-390002, India
[2]Institute of Theoretical Science, University of Oregon, Eugene, OR 97403-1258, U.S.A.

**Abstract**

We have performed a numerical analysis of branching ratios and direct CP-asymmetries of B meson decays to two light pseudoscalar nonets with $|\Delta S| = 1$ using QCD improved factorization. The parameters related to B meson including its semileptonic form factors, strange quark mass and the CKM phase angle have been varied over a limited range and phenomenological parameters used to parameterize divergences in hard spectator scattering and weak annihilation have been varied over a wide range simultaneously with an aim to achieve a fit of branching ratios and some direct CP-asymmetries. A good fit for the majority of the parameters has been achieved indicating that the standard model is quite satisfactory in accounting for these processes.

**Introduction**

Nonleptonic two-body decays of B mesons are extremely interesting since they provide valuable tests of the standard model and may reveal existence of physics beyond standard model (SM) if theoretical analysis based on SM fails to achieve a satisfactory explanation of the experimental results. In particular, many of these decay modes carry interesting information on CP-violating interactions and on flavor parameters of the SM. Among the rare nonleptonic B decays, a large disparity observed between branching ratios (BRs) to $\eta' K$ and $\eta K$ decays while BRs to $\pi K$ decays being intermediate between these two decays have puzzled the community for last one and a half decades [1,2]. The modes with a single kaon based on $|\Delta S| = 1$ transition have "large" BRs of the order of few times $10^{-6}$ to few times $10^{-5}$[2]. An interesting aspect of charmless B decays is that it involves three distinct scales: $M_W \gg m_b \gg \Lambda_{QCD}$. While sensitivity to the weak interaction scale and probable new phenomena at this scale is most interesting, the most theoretical work is related to strong interaction corrections. The mass scale $\sim m_b$ arises naturally due to the energy available to decay products in the B rest frame and also as the typical parton off-shellness in some of the loop diagrams. The process of separation of physics of different scales from each other, is generally called "factorization". The naive factorization approach [3], adopted initially for calculating these decays, have been superseded by QCD improved factorization[4-12] and perturbativeQCD[13-16] calculations. QCD factorization can be used to compute two-body decay amplitudes from first principle. Its accuracy is limited only by power corrections to the heavy quark limit and the uncertainties of theoretical inputs such as quark masses, CKM matrix elements, form factors, light-cone distribution amplitudes, mesonic mixing parameters, etc.

In QCD factorization approach, strong final-state interaction phases arise from the hard scattering kernel and are, therefore, calculable in the heavy quark limit. These strong phases are very important for studying CP-violation in B physics. The calculation of the amplitudes for B→$P_1P_2$ decays involves the twist-three light-cone wave function of the light pseudoscalar. Unfortunately, this gives rise to a logarithmic divergence in the contribution from hard spectator scattering at the endpoint of the twist-three light-cone distribution amplitude signaling breakdown of QCD factorization at twist-three level. BBNS [5] have also taken into account contribution from weak annihilation, which is power suppressed within the framework of QCD factorization. The amplitude of weak annihilation also contains endpoint divergences indicating existence of soft contributions. BBNS have given a phenomenological treatment of the endpoint divergences in both the hard spectator scattering, called $X_H$ and the weak annihilation, called $X_A$. Since the predictions of the BRs for B→$P_1P_2$ decays strongly depend on the parameters $X_H$ and $X_A$, which are treated as universal for these decays, it is essential to reliably estimate the effects of $X_H$ and $X_A$. In this work we study the decay processes $B^{\pm(0)} \to \eta^{(\prime)} K^{\pm(0)}, \pi^0 K^{\pm(0)}, \pi^{\pm} K^0, \pi^{(\pm)} K^{(\mp)}$ using QCD factorization. We study their branching ratios as well as the direct CP-asymmetries, for the cases where it has been observed with accuracy better than $2\sigma$, simultaneously while varying strange quark mass, the CKM phase angle $\gamma$, hadronic transition form factors and phenomenological parameters $X_H$ and $X_A$ over specified ranges so as to get the best fit.

For accommodating dramatically different data for B→ $\eta K$ and B→ $\eta' K$ BRs within perturbative QCD and QCD factorization different proposals have been tried: FKS scheme[17] for $\eta - \eta'$ mixing with a significant flavor singlet contribution [8], a large B→ $\eta'$ transition form factor[12], a high chiral scale $m_0^q$ associated with $\eta_q$ meson [18], an enhanced hadronic matrix element $\langle 0|\bar{s}i\gamma_5 s| \eta' \rangle$ [19], the long distance charming penguin and gluonic charming penguin in SCET [20], inelastic final state interaction (FSI) [21], $\eta - \eta' - G$ mixing scheme and $\eta - \eta' - G - \eta_c$ mixing scheme [22]. Usually, the theoretical errors in BRs and CP-asymmetries arising due to uncertainties in each of the input parameters are quoted separately around a central value [6,8,9,16,20], or the errors are added in quadrature[8,16] and the result displayed. In some cases ratios of branching fractions are exhibited to cancel some common errors [5,6,9]. In all such analyses one is never sure as to which combination of input parameters will yield a result which will match with the corresponding experimental result or will be within a certain close range to the experimental result. We feel it worthwhile to search a set of input parameters for which QCD factorization gives results that agree with the experimental results within the experimental uncertainties or are close to the experimental results for as many BRs and CP-asymmetries as possible. This kind of analysis will constrain the ranges of input parameters, may suggest incorporation of higher order terms not included so far in theoretical calculations or, depending upon the amount of discrepancies, may also suggest existence of new physics beyond SM. With this in mind, we will do a rigorous numerical analysis of all the BRs and CP-asymmetries mentioned above using FKS mixing scheme [17].

**Parameterization of B decays to two light pseudoscalars**

Using operator product expansion, the relevant effective Hamiltonian for $|\Delta B| = 1$ decays is given by[23]:

$$\mathcal{H}_{eff} = \frac{G_F}{\sqrt{2}}\left\{\sum_{p=u,c} V_{pb}V_{pq}^*\left(C_1 O_1^p + C_2 O_2^p + \sum_{i=3}^{10} C_i O_i\right) - V_{tb}V_{tq}^*\left(C_{7\gamma}O_{7\gamma} + C_{8g}O_{8g}\right)\right\} + H.C.,$$

(q=d,s)  (1)

where $O_{1,2}^p$ are the left-handed current-current operators arising from W-boson exchange, $O_{3,...6}$ and $O_{7,...10}$ are QCD and electroweak penguin operators, and $O_{7\gamma}$ and $O_{8g}$ are the electromagnetic and chromomagnetic dipole operators. The Wilson coefficients (WCs) $C_i(\mu)$ are obtained at a high scale $\mu \sim M_W$ and evolved down to a characteristic scale $\mu \sim m_b$ using next-to-leading logarithmic order in the NDR scheme. Their numerical values have been taken from Ref. [5].

In the QCDF approach in the heavy quark limit, the hadronic matrix element for B→$P_1P_2$ due to a given operator $O_i$ can be written as

$$\langle P_1 P_2 | O_i | B \rangle = \langle P_1 P_2 | O_i | B \rangle_{NF} \left[1 + \sum_n r_n(\alpha_s)^n + O\left(\frac{\Lambda_{QCD}}{m_b}\right)\right] \quad (2)$$

In Eq.(2), $\langle P_1 P_2 | O_i | B \rangle_{NF}$ denotes the naïve factorization result while the second and third term in the square bracket denote the radiative correction in $\alpha_s$ and the power corrections in $\Lambda_{QCD}/m_b$.

The decay amplitude for B→$P_1P_2$ can be written as

$$\mathcal{A}(B \to P_1 P_2) = \frac{G_F}{\sqrt{2}} \sum_{p=u,c} \sum_{i=1}^{10} V_{pb}V_{pq}^*\left(a_i^p \langle P_1 P_2 | O_i | B \rangle_{NF} + f_B f_{P_1} f_{P_2} b_i\right) \quad (3)$$

Here the first term on the r.h.s. includes vertex corrections, penguin corrections and hard spectator scattering contributions which are absorbed into the QCD coefficients $a_i^p$, whereas the second term includes weak annihilation contributions which are absorbed into the parameters $b_i$. We list below the relevant decay amplitudes[5,7,9,11]:

$$\mathcal{A}(B^- \to \eta^{(\prime)} K^-) = -i\frac{G_F}{\sqrt{2}} f_K F_0^{B \to \eta^{(\prime)}}(m_K^2)(m_B^2 - m_{\eta^{(\prime)}}^2)\left[V_{ub}V_{us}^*(a_1' + a_4^{u\prime} + a_{10}^{u\prime} + (a_6^{u\prime} + a_8^{u\prime})R_1) + V_{cb}V_{cs}^*(a_4^{c\prime} + a_{10}^{c\prime} + (a_6^{c\prime} + a_8^{c\prime})R_1)\right] - i\frac{G_F}{\sqrt{2}} F_0^{B \to K}(m_{\eta^{(\prime)}}^2)(m_B^2 - m_K^2)\left\{f_{\eta^{(\prime)}}^u \left[V_{ub}V_{us}^*\left(a_2 + 2a_3 - 2a_5 - \frac{1}{2}(a_7 - a_9) - \left(a_6^u - \frac{1}{2}a_8^u\right)R_3\right) + V_{cb}V_{cs}^*\left(2a_3 - 2a_5 - \frac{1}{2}(a_7 - a_9) - \left(a_6^c - \frac{1}{2}a_8^c\right)R_3\right)\right] + f_{\eta^{(\prime)}}^s \left[V_{ub}V_{us}^*\left(a_3 + a_4^u - a_5 + \frac{1}{2}(a_7 - a_9 - a_{10}^u) + \left(a_6^u - \frac{1}{2}a_8^u\right)R_3\right) + V_{cb}V_{cs}^*\left(a_3 + a_4^c - a_5 + \frac{1}{2}(a_7 - a_9 - a_{10}^c) + \left(a_6^c - \frac{1}{2}a_8^c\right)R_3\right)\right]\right\} - i\frac{G_F}{\sqrt{2}} f_B f_K (f_{\eta^{(\prime)}}^u + f_{\eta^{(\prime)}}^s)[V_{ub}V_{us}^* b_2 + (V_{ub}V_{us}^* + V_{cb}V_{cs}^*)(b_3 + b_3^{ew})], \quad (4)$$

$$\mathcal{A}(\bar{B}^0 \to \eta^{(\prime)} \bar{K}^0) = -i\frac{G_F}{\sqrt{2}} f_K F_0^{B \to \eta^{(\prime)}}(m_K^2)(m_B^2 - m_{\eta^{(\prime)}}^2)\left[V_{ub}V_{us}^*\left(a_4^{u\prime} - \frac{1}{2}a_{10}^{u\prime} + \left(a_6^{u\prime} - \frac{1}{2}a_8^{u\prime}\right)R_2\right) + V_{cb}V_{cs}^*\left(a_4^{c\prime} - \frac{1}{2}a_{10}^{c\prime} + \left(a_6^{c\prime} - \frac{1}{2}a_8^{c\prime}\right)R_2\right)\right] -$$

$$i\frac{G_F}{\sqrt{2}}F_0^{B\to K}\left(m_{\eta^{(\prime)}}^2\right)(m_B^2 - m_K^2)\left\{f_{\eta^{(\prime)}}^u\left[V_{ub}V_{us}^*\left(a_2 + 2a_3 - 2a_5 - \frac{1}{2}(a_7 - a_9) - \left(a_6^u - \frac{1}{2}a_8^u\right)R_3\right) + V_{cb}V_{cs}^*\left(2a_3 - 2a_5 - \frac{1}{2}(a_7 - a_9) - \left(a_6^c - \frac{1}{2}a_8^c\right)R_3\right)\right] + f_{\eta^{(\prime)}}^s\left[V_{ub}V_{us}^*\left(a_3 + a_4^u - a_5 + \frac{1}{2}(a_7 - a_9 - a_{10}^u) + \left(a_6^u - \frac{1}{2}a_8^u\right)R_3\right) + V_{cb}V_{cs}^*\left(a_3 + a_4^c - a_5 + \frac{1}{2}(a_7 - a_9 - a_{10}^c) + \left(a_6^c - \frac{1}{2}a_8^c\right)R_3\right)\right]\right\} - i\frac{G_F}{\sqrt{2}}f_B f_K(f_{\eta^{(\prime)}}^u + f_{\eta^{(\prime)}}^s)(V_{ub}V_{us}^* + V_{cb}V_{cs}^*)(b_3 - \frac{1}{2}b_3^{ew}),$$  (5)

$$\mathcal{A}(B^- \to \pi^- \overline{K}^0) = i\frac{G_F}{\sqrt{2}}f_K F_0^{B\to\pi}(m_K^2)(m_B^2 - m_\pi^2)\left[V_{ub}V_{us}^*\left(a_4^u - \frac{1}{2}a_{10}^u + \left(a_6^u - \frac{1}{2}a_8^u\right)r_\chi^K\right) + V_{cb}V_{cs}^*\left(a_4^c - \frac{1}{2}a_{10}^c + \left(a_6^c - \frac{1}{2}a_8^c\right)r_\chi^K\right)\right] + i\frac{G_F}{\sqrt{2}}f_B f_K f_\pi[V_{ub}V_{us}^*(b_2 + b_3 + b_3^{ew}) + V_{cb}V_{cs}^*(b_3 + b_3^{ew})],$$  (6)

$$-\sqrt{2}\mathcal{A}(B^- \to \pi^0 K^-) = i\frac{G_F}{\sqrt{2}}f_K F_0^{B\to\pi}(m_K^2)(m_B^2 - m_\pi^2)\left[V_{ub}V_{us}^*\left(a_1 + a_4^u + a_{10}^u + (a_6^u + a_8^u)r_\chi^K\right) + V_{cb}V_{cs}^*\left(a_4^c + a_{10}^c + (a_6^c + a_8^c)r_\chi^K\right)\right] + i\frac{G_F}{\sqrt{2}}f_\pi F_0^{B\to K}(m_\pi^2)(m_B^2 - m_K^2)\left[V_{ub}V_{us}^*\left(a_2 + \frac{3}{2}(-a_7 + a_9)\right) + V_{cb}V_{cs}^*\frac{3}{2}(-a_7 + a_9)\right] + i\frac{G_F}{\sqrt{2}}f_B f_K f_\pi[V_{ub}V_{us}^* b_2 + (V_{ub}V_{us}^* + V_{cb}V_{cs}^*)(b_3 + b_3^{ew})],$$  (7)

$$-\mathcal{A}(\overline{B}^0 \to \pi^+ K^-) = = i\frac{G_F}{\sqrt{2}}f_K F_0^{B\to\pi}(m_K^2)(m_B^2 - m_\pi^2)\left[V_{ub}V_{us}^*\left(a_1 + a_4^u + a_{10}^u + (a_6^u + a_8^u)r_\chi^K\right) + V_{cb}V_{cs}^*\left(a_4^c + a_{10}^c + (a_6^c + a_8^c)r_\chi^K\right)\right] + i\frac{G_F}{\sqrt{2}}f_B f_K f_\pi(V_{ub}V_{us}^* + V_{cb}V_{cs}^*)(b_3 - \frac{1}{2}b_3^{ew}),$$  (8)

and from SU(2) sum rule, we have [20]

$$\sqrt{2}\mathcal{A}(\overline{B}^0 \to \pi^0 \overline{K}^0) = -\mathcal{A}(B^- \to \pi^- \overline{K}^0) + \sqrt{2}\mathcal{A}(B^- \to \pi^0 K^-) - \mathcal{A}(\overline{B}^0 \to \pi^+ K^-),$$  (9)

where

$$R_{1(2)} = \frac{2m_{K^{(0)}}^2}{(m_b - m_q)(m_q + m_s)}, \quad R_3 = \frac{m_{\eta^{(\prime)}}^2}{m_s(m_b - m_s)}, \quad r_\chi^K = \frac{2m_K^2}{m_b(m_q + m_s)}, \quad r_\chi^\pi = \frac{m_\pi^2}{m_b m_q},$$  (10)

The coefficients $a_i^{(\prime)}$ and $b_i$ are expressed as

$$a_1^{(\prime)} = C_1 + \frac{C_2}{3}\left[1 + \frac{C_F \alpha_s}{4\pi}\left(V_P + \frac{4\pi^2}{3}H(BP_1, P_2)\right)\right],$$

$$a_2 = C_2 + \frac{C_1}{3}\left[1 + \frac{C_F \alpha_s}{4\pi}\left(V_P + \frac{4\pi^2}{3}H(BP_1, P_2)\right)\right],$$

$$a_3 = C_3 + \frac{C_4}{3}\left[1 + \frac{C_F \alpha_s}{4\pi}\left(V_P + \frac{4\pi^2}{3}H(BP_1, P_2)\right)\right],$$

$$a_4^{p(\prime)} = C_4 + \frac{C_3}{3}\left[1 + \frac{C_F \alpha_s}{4\pi}\left(V_P + \frac{4\pi^2}{3}H(BP_1, P_2)\right)\right] + \frac{C_F \alpha_s}{12\pi}P_{P,2}^p,$$

$$a_5 = C_5 + \frac{C_6}{3}\left[1 - \frac{C_F \alpha_s}{4\pi}\left(V_P + 12 + \frac{4\pi^2}{3}H(BP_1, P_2)\right)\right],$$

$$a_6^{p(')} = C_6 + \frac{C_5}{3}\left[1 - 6\frac{C_F \alpha_s}{4\pi}\right] + \frac{C_F \alpha_s}{12\pi}P_{P,3}^p,$$

$$a_7 = C_7 + \frac{C_8}{3}\left[1 - \frac{C_F \alpha_s}{4\pi}\left(V_P + 12 + \frac{4\pi^2}{3}H(BP_1, P_2)\right)\right],$$

$$a_8^{p(')} = C_8 + \frac{C_7}{3}\left[1 - 6\frac{C_F \alpha_s}{4\pi}\right] + \frac{\alpha}{27\pi}P_{P,3}^{p,ew},$$

$$a_9 = C_9 + \frac{C_{10}}{3}\left[1 + \frac{C_F \alpha_s}{4\pi}\left(V_P + \frac{4\pi^2}{3}H(BP_1, P_2)\right)\right],$$

$$a_{10}^{p(')} = C_{10} + \frac{C_9}{3}\left[1 + \frac{C_F \alpha_s}{4\pi}\left(V_P + \frac{4\pi^2}{3}H(BP_1, P_2)\right)\right] + \frac{\alpha}{27\pi}P_{P,2}^{p,ew},$$

$$b_2 = \frac{C_F}{9}C_2 A^i, \quad b_3 = \frac{C_F}{9}[C_3 A^i + A^f(C_5 + 3C_6)], \quad b_3^{ew} = \frac{C_F}{9}[C_9 A^i + A^f(C_7 + 3C_8)], \quad (11)$$

where $C_F = \frac{4}{3}$ is the color factor ($N_C = 3$), the superscript p=u,c, $C_i = C_i(\mu)$ and $\alpha_s = \alpha_s(\mu)$. The parameters $V_P$ resulting from the vertex corrections, $P_{P,2}^p$ and $P_{P,3}^p$ ($P_{P,2}^{p,ew}$) arising from QCD (electroweak) penguin contractions and the contributions from the dipole operators, and $H(BP_1, P_2)$ originating from hard spectator scattering are given as [5,7,9]:

$$V_P = 12 \ln\frac{m_b}{\mu} - 18 + \int_0^1 dx\, g(x)\Phi_P(x),$$

$$H(BP_1, P_2) = \frac{f_B f_{P_1}}{m_B^2 F_0^{B \to P_1}} \int_0^1 dz\, \frac{\Phi_B(z)}{z} \int_0^1 dx\, \frac{\Phi_{P_2}(x)}{1-x} \int_0^1 dy\, \left[\frac{\Phi_{P_1}(y)}{1-y} + \frac{2\mu_{P_1}}{m_b}\frac{(1-x)}{x}\frac{\Phi_{P_1}^p(y)}{1-y}\right],$$

$$A^i \approx \pi\alpha_s\left[18\left(X_A - 4 + \frac{\pi^2}{3}\right) + 2r_\chi^K r_\chi^P X_A^2\right],$$

$$A^f \approx 6\pi\alpha_s(r_\chi^K + r_\chi^P)X_A(2X_A - 1), \quad (12)$$

where

$$g(x) = 3\left(\frac{1-2x}{1-x}\ln x - i\pi\right) + \left[2Li_2(x) - \ln^2 x + \frac{2\ln x}{1-x} - (3 + 2i\pi)\ln x - (x \leftrightarrow 1 - x)\right]. \quad (13)$$

For the wave function $\Phi_B(z)$ of the B meson, the following parameterization has been used:

$$\int_0^1 dx\, \frac{\Phi_B(x)}{x} = \frac{m_B}{\lambda_B}, \quad (14)$$

where for $\lambda_B = 350 \pm 150\ MeV$ has been suggested [5,9]. For light pseudoscalar nonets, the asymptotic forms of respective LCDA has been used for the leading twist and twist-3 LCDAs [5]:

$$\Phi_P(x) = 6x(1-x),$$

$$\Phi_P^p(x) = 1. \quad (15)$$

We also have

$$\mu_P = \frac{m_P^2}{m_1+m_2} \text{ for P} = \pi \text{ and K} \tag{16a}$$

($m_1$ and $m_2$ are current quark masses of the valence quarks of the meson P), while chirally enhanced parameter $r_\chi = \frac{2\mu_P}{m_b}$. For $r_\chi^{\eta^{(\prime)}}$ we take[7]:

$$r_\chi^{\eta^{(\prime)}}\left(1 - \frac{f_{\eta^{(\prime)}}^u}{f_{\eta^{(\prime)}}^s}\right) = r_\chi^\pi = r_\chi^K \equiv r_\chi. \tag{16b}$$

The phenomenological parameters for the end-point divergent integrals [5]

$$X_{H,A} \equiv \int_0^1 \frac{dx}{x} \equiv \left(1 + \rho_{H,A} e^{i\phi_{H,A}}\right) \ln\frac{m_B}{\Lambda_h}, \tag{17}$$

have been introduced with $X_H$ and $X_A$ for hard spectator scattering contribution and the annihilation contribution respectively. Here the phases $\phi_{H,A}$ are arbitrary $0 \leq \phi_{H,A} \leq 2\pi$, and the parameters $\rho_{H,A} \leq 2$ have been taken while the phenomenological scale $\Lambda_h = 0.5\ GeV$ has been assumed [5]. The coefficients $a_i$ and $a_i'$ in Eq.(11) include different vertex and hard spectator scattering contributions as follows : for $a_i$, $V_P = V_{\eta^{(\prime)}}$ or $V_\pi$ and $H(BP_1, P_2) = H(B\eta^{(\prime)}, K)$ or $H(B\pi, K)$ while for $a_i'$, $V_P = V_K$ and $H(BP_1, P_2) = H(BK, \eta^{(\prime)})$. For the expressions of the QCD penguin parameters $P_{P,i}^p$ and the electroweak penguin parameters $P_{P,i}^{p,ew}$, we refer readers to Refs.[5,9].

The following constants have been used in this work (masses and decay constants in GeV unit) [24,25,9,26,27] :

$f_B = 0.2 \pm 0.01$, $f_K = 0.16$, $\Lambda_{\overline{MS}}^{(5)} = 0.225$, $\Lambda_{\overline{MS}}^{(4)} = 0.319$ (both mass parameters to two-loop accuracy), $m_q = 0.0041$, $m_s = 0.1 \pm 0.01$, $m_c = 1.4$, $m_b = 4.2$, $f_\eta^u = 0.0778$, $f_\eta^s = -0.1112$, $f_{\eta'}^u = 0.0636$, $f_{\eta'}^s = 0.1413$, $m_B = 5.28$, $f_\pi = 0.131$, $V_{us} = 0.2246$, $|V_{ub}| = 0.00361$, $V_{cs} = 0.9748$, $V_{cb} = 0.04197$, $\gamma = 60° \pm 5°$, $\tau_{B^0} = 1.52 \times 10^{-12} sec$, $\tau_{B^\pm} = 1.641 \times 10^{-12} sec$, $\alpha = 1/129$. (18)

The numerical values of $f_{\eta^{(\prime)}}^{u,s}$ are in the standard FKS scheme [17]. In evaluation of $a$ and $b$ parameters, $C_i$ and $\alpha_s$ have been evaluated at the renormalization scale $\mu = m_b$; however, when these parameters appear as coefficients of $H(BP_1, P_2)$, $\mu = \mu_h = \sqrt{\Lambda_h m_b}$ with $\Lambda_h = 0.5$ GeV has been used [5]. Lattice calculations give $f_B = 190.6(4.7)\ MeV$ [26] while Borel QCD sum rule give $\overline{m}_b(\overline{m}_b) = 4.247 \pm 0.027 \pm 0.011$ GeV [27]. While the large quark masses are defined at $\mu = m_b$, the small quark masses are defined at $\mu = 2\ GeV$[9]. For semileptonic form factors $F_0^{B \to \eta^{(\prime)}}(q^2)$, we have used the parameterization [28]:

$$F_0^{B \to \eta^{(\prime)}}(0) = F_1 \frac{f_{\eta^{(\prime)}}^u}{f_\pi} + F_2 \frac{2f_{\eta^{(\prime)}}^u + f_{\eta^{(\prime)}}^s}{\sqrt{6}f_\pi}. \tag{19}$$

To account for small variation in the form factors for $q^2 = m_P^2$, we have used the parameterization of Ball and Zwicky [29]:
$F_0^{B \to P}(q^2) = \frac{r_2}{1-q^2/m_{fit}^2}$, where $m_{fit}^2 = 37.46 \ GeV^2$ for P=K and $m_{fit}^2 = 31.03 \ GeV^2$ for P=$\eta^{(\prime)}$.
$F_0^{B \to P}(0) = r_2$ and we assume same $m_{fit}^2$ for $\eta'$ as for $\eta$.

**Numerical analysis and results**

For numerical analysis, we have varied $F_1$, $F_2$, $m_s$, $\gamma$, $F_0^{B \to \pi}$, $F_0^{B \to K}$, $\lambda_B$ and $f_B$ over a limited range (around 10%), since their values used in the current literature have some uncertainties. The phenomenological parameters $\rho_{H,A}$ and $\phi_{H,A}$ have been varied over a wide range: $0 \leq \rho_{H,A} \leq 2$ and $0 \leq \phi_{H,A} \leq 2\pi$. All the 12 parameters listed in Table I were varied simultaneously to get a good fit for 7 BRs and 1 CP-asymmetry listed in the first eight columns of Table II. The quality of fit was judged by the parameters $\chi$ and $\chi_1$ defined as follows: $\chi^2 = \frac{1}{N} \sum_{i=1}^{N} \frac{(f_i - f_i^e)^2}{(f_i + f_i^e)^2}$, where $f_i$ is our computed result and $f_i^e$ is the nearest edge value of the measured result arising due to the error bar and $f_i - f_i^e = 0$ if $f_i$ lies within the range of errors of the measured result. $\chi_1^2 = \frac{1}{N} \sum_{i=1}^{N} \frac{(f_i - f_i^c)^2}{(f_i + f_i^c)^2}$, where $f_i$ is our computed result and $f_i^c$ is the central value of the measured result and $f_i - f_i^c = 0$ if $f_i$ lies within the range of errors of the measured result. The nomenclatures $\chi$ and $\chi_1$ have been used for the first 8 output parameters listed in Tables II and IV and nomenclatures $\chi'$ and $\chi_1'$ have been used for the first 11 output parameters listed there. The values of $\chi$, $\chi_1$, $\chi'$ and $\chi_1'$ have been listed in the last four columns of Tables II and IV. For comparison, the experimental results for the 12 output parameters have been listed in the second row of Tables II and IV. Two sets of results have been obtained in Table II for $\chi < 1.5, 2.1$. Both the input parameters listed in Table I and the resulting output parameters listed in Table II appear in a range while satisfying the condition $\chi < 1.5, 2.1$. It has been argued that the largest uncertainty in BRs and CP-asymmetries come from the phenomenological parameters $\rho_{H,A}$ and $\phi_{H,A}$ [5,9]. We have put an upper limit on $\rho_{H,A}$: $\rho_{H,A} \leq 2$. Normally, $\rho_{H,A} \leq 1$ has been used for analysis of B decays [5-12]. However, a conservative estimate of $\rho_{H,A} \leq 2$ has also been advocated [7], $\rho_A = 2.0$ has been used [9] and even $\rho_H \approx 4.9$ has been suggested to solve the B→K$\pi$ CP puzzle [30]. We observe that the values of $\rho_H$ required for a good fit is always on the higher side, close to the upper limit which we have set. In contrast to this, $\rho_A$ lies in an intermediate region. The values of $\phi_{H,A}$ in a specific range, as shown in Table I, give a good fit. As stated earlier, the values of the remaining 8 input parameters have been varied in a range which has been used by several authors in recent years. Hence, it is gratifying to find that an appropriate choice of the 4 parameters $\rho_{H,A}$ and $\phi_{H,A}$ can produce the experimental results for 7 BRs and a CP-asymmetry so closely. In Table IV we have listed a few typical results with low $\chi$'s for the input parameters listed in Table III. To illustrate the fact that larger values of $\rho_H$ further reduces $\chi$ and $\chi_1$, in Table IV, at the bottom, we have displayed a typical result for $\rho_H$=2.2. We have also computed $Br(\bar{B}^0 \to \bar{K}^0 \pi^0)$, $A_{CP}(B^- \to K^- \eta')$ and $A_{CP}(B^- \to K^- \pi^0)$, and the corresponding $\chi'$ and $\chi_1'$ for all the first 11 output parameters shown in Table II for the same set of input parameters which yield $\chi < 1.5, 2.1$ and listed there. In Table IV also we have listed the 11 output parameters and $\chi'$ and $\chi'_1$ for the input parameters shown in Table III. For the sake of comparison, we have also listed the resulting values of $A_{CP}(\bar{B}^0 \to K^-\pi^+)$ in Table II as well as Table IV, although it has not been used in the analysis. The $Br(\bar{B}^0 \to \bar{K}^0\pi^0)$, $A_{CP}(B^- \to K^-\eta')$ and $A_{CP}(B^- \to K^-\pi^0)$, depend on amplitudes which have already been used to minimize $\chi$ and $\chi_1$. The fact that the values of $\chi'$ and $\chi'_1$ are significantly higher than $\chi$ and $\chi_1$ indicates that the strong interaction phases have not been accounted for sufficiently well in the present approach. For $A_{CP}(\bar{B}^0 \to K^-\pi^+)$, the fit does not account for even the sign correctly. The latter observation has been noted by other authors too [9,30]. It may be amusing to note that our result for $\Delta_{K\pi} = A_{CP}(B^- \to K^-\pi^0) - A_{CP}(\bar{B}^0 \to K^-\pi^+) \approx 10\%$ has approximately correct value. There are a couple of suggestions to resolve discrepancies found in observables related to $\bar{B} \to \bar{K}\pi$ decays within QCDF. One way may be to take a large $\rho_H \approx 4.9$ in calculation of the coefficient $a_2$ [30]. Another approach may be to include color allowed decay $B^- \to K^-\eta'$ (which has a large BR) followed by the rescattering of $K^-\eta' \to K^-\pi^0$ [31].

TABLE I: The input parameters, shown in the first row, have been varied in a range shown in next two rows (except for the $\gamma$ in the second row having a single value). The corresponding output parameters have been shown in TABLE II.

| $F_1$ | $F_2$ | $m_s$ (GeV) | $\gamma$ | $F_0^{B\to\pi}$ | $\lambda_B$ (GeV) | $f_B$ (GeV) | $F_0^{B\to K}$ | $\rho_H$ | $\phi_H/\pi$ | $\rho_A$ | $\phi_A/\pi$ |
|---|---|---|---|---|---|---|---|---|---|---|---|
| 0.248-0.265 | 0.003-0.004 | 0.09-0.1 | 55° | 0.23-0.26 | 0.21-0.24 | 0.19-0.21 | 0.325-0.34 | 1.92-2.0 | 1.45-1.50 | 1.27-1.43 | 0.14-0.165 |
| 0.248-0.265 | 0.003-0.004 | 0.09-0.11 | 55° $-65°$ | 0.23-0.26 | 0.21-0.24 | 0.19-0.21 | 0.325-0.355 | 1.92-2.0 | 1.44-1.50 | 1.27-1.45 | 0.09-0.165 |

TABLE II: Summary of results of our numerical analysis: The second row shows the experimental results for the branching ratios and direct CP-asymmetries shown in the first row. The third and the fourth rows are the computed results for the quantities shown in the first row corresponding to the input parameters shown in the second and the third rows respectively of TABLE I. The output parameters take values in a range because input parameters have been varied in a range. The $\chi$'s have been defined in the text.

| $Br(B^- \to K^-\eta) \times 10^6$ | $A_{CP}(B^- \to K^-\eta) \times (-1)$ (%) | $Br(B^- \to K^-\eta') \times 10^6$ | $Br(B^- \to K^-\pi^0) \times 10^6$ | $Br(\bar{B}^0 \to \bar{K}^0\eta) \times 10^6$ | $Br(\bar{B}^0 \to \bar{K}^0\eta') \times 10^6$ | $Br(B^- \to \bar{K}^0\pi^-) \times 10^6$ | $Br(\bar{B}^0 \to K^-\pi^+) \times 10^6$ | $Br(\bar{B}^0 \to \bar{K}^0\pi^0) \times 10^6$ | $A_{CP}(B^- \to K^-\eta')$ (%) | $A_{CP}(B^- \to K^-\pi^0)$ (%) | $A_{CP}(\bar{B}^0 \to K^-\pi^+)$ (%) | $\chi$ (%) | $\chi_1$ (%) | $\chi'$ (%) | $\chi'_1$ (%) |
|---|---|---|---|---|---|---|---|---|---|---|---|---|---|---|---|
| $2.36^{+0.22}_{-0.21}$ | $37\pm8$ | $71.1\pm2.6$ | $12.1\pm0.8$ | $1.23^{+0.27}_{-0.24}$ | $66.1\pm3.1$ | $24.1\pm1.8$ | $18.9\pm0.7$ | $10.1\pm1.0$ | $1.3\pm1.7$ | $5.1\pm2.5$ | $-10.7^{+2.2}_{-1.2}$ | | | | |
| 2.04-2.14 | 27.53-29.17 | 66.68-73.91 | 13.46-13.70 | 1.47-1.55 | 63.43-69.51 | 21.40-22.13 | 17.82-18.23 | 6.75-7.07 | 1.93-2.82 | 18.42-20.10 | 8.82-10.57 | <1.5 | 3.56-7.23 | 12.76-13.81 | 19.32-21.89 |
| 2.00-2.14 | 25.02-29.17 | 66.28-73.96 | 13.17-13.70 | 1.45-1.58 | 62.06-69.55 | 20.73-22.16 | 17.44-18.25 | 6.37-7.07 | 1.09-2.82 | 14.12-20.10 | 5.02-10.57 | <2.1 | 3.56-7.80 | 9.59-13.81 | 15.87-21.89 |

In Table II, the outputs have been listed over a range since the corresponding inputs in Table I have been applied over a range. On the other hand, in Table IV, few typical results with low $\chi$ values have been listed corresponding to single inputs listed in Table III. It is observed that low $\chi$ and $\chi_1$ does not imply low $\chi'$ and $\chi'_1$ and vice versa. It is also seen that low $F_1$, $m_s$ and $\gamma$ favor low $\chi$ and $\chi_1$, whereas the higher values of these parameters favor low $\chi'$ and $\chi'_1$. It is also observed that the BRs of B decays involving $\eta'$ are obtained in a range which well covers the range of the corresponding experimental results within the error bars. For the remaining BRs, the computed results just touch the boundaries of the experimental errors from one side and they mostly lie outside the range of the error bars. In this work, we have not varied $\eta - \eta'$ mixing parameters which have some uncertainties [32]. There are works in which mixing of $\eta - \eta'$ with glueball states [33] and also with glueball as well as $\eta_c$ is considered [22]. This may make some improvement in results on BRs and direct CP-asymmetries of decays involving $\eta$ and $\eta'$ mesons.

TABLE III: The input parameters, shown in the first row, and their values shown in next four rows. The corresponding output parameters have been shown in TABLE IV.

| $F_1$ | $F_2$ | $m_s$ (GeV) | $\gamma$ | $F_0^{B\to\pi}$ | $\lambda_B$ (GeV) | $f_B$ (GeV) | $F_0^{B\to K}$ | $\rho_H$ | $\phi_H/\pi$ | $\rho_A$ | $\phi_A/\pi$ |
|---|---|---|---|---|---|---|---|---|---|---|---|
| 0.248 | 0.0034 | 0.09 | 55° | 0.24 | 0.21 | 0.21 | 0.325 | 2.0 | 1.45 | 1.3 | 0.155 |
| 0.265 | 0.0034 | 0.09 | 55° | 0.24 | 0.23 | 0.21 | 0.335 | 1.97 | 1.43 | 1.34 | 0.125 |
| 0.26 | 0.0034 | 0.11 | 60° | 0.25 | 0.21 | 0.2 | 0.345 | 1.94 | 1.46 | 1.38 | 0.09 |
| 0.265 | 0.0034 | 0.09 | 55° | 0.24 | 0.23 | 0.2 | 0.335 | 2.2 | 1.46 | 1.325 | 0.12 |

TABLE IV: : Some typical results with low $\chi$'s :The second row shows the experimental results for the branching ratios and direct CP-asymmetries shown in the first row. The third through the sixth rows are the computed results for the quantities shown in the first row corresponding to the input parameters shown in the second through the fifth rows respectively of TABLE III. The low $\chi$ values have been shown in bold digits.

| $Br(B^-\to K^-\eta)\times 10^6$ | $A_{CP}(B^-\to K^-\eta)\times(-1)$ (%) | $Br(B^-\to K^-\eta')\times 10^6$ | $Br(B^-\to K^-\pi^0)\times 10^6$ | $Br(\bar{B}^0\to \bar{K}^0\eta)\times 10^6$ | $Br(\bar{B}^0\to \bar{K}^0\eta')\times 10^6$ | $Br(B^-\to \bar{K}^0\pi^-)\times 10^6$ | $Br(\bar{B}^0\to K^-\pi^+)\times 10^6$ | $Br(\bar{B}^0\to \bar{K}^0\pi^0)\times 10^6$ | $A_{CP}(B^-\to K^-\eta')$ (%) | $A_{CP}(B^-\to K^-\pi^0)$ (%) | $A_{CP}(\bar{B}^0\to K^-\pi^+)$ (%) | $\chi$ (%) | $\chi_1$ (%) | $\chi'$ (%) | $\chi'_1$ (%) |
|---|---|---|---|---|---|---|---|---|---|---|---|---|---|---|---|
| $2.36^{+0.22}_{-0.21}$ | 37±8 | 71.1 ±2.6 | 12.1 ±0.8 | $1.23^{+0.27}_{-0.24}$ | 66.1 ±3.1 | 24.1 ±1.8 | 18.9 ±0.7 | 10.1 ±1.0 | 1.3 ±1.7 | 5.1 ±2.5 | $-10.7^{+2.2}_{-1.2}$ | | | | |
| 2.11 | 28.70 | 69.95 | 13.54 | 1.51 | 66.00 | 21.67 | 17.90 | 6.84 | 2.25 | 19.56 | 9.32 | **1.12** | 6.73 | 13.47 | 20.45 |
| 2.10 | 29.01 | 68.52 | 13.57 | 1.47 | 65.06 | 21.61 | 18.00 | 6.82 | 2.94 | 19.38 | 10.60 | 1.14 | **3.56** | 13.36 | 21.48 |
| 2.02 | 25.02 | 72.44 | 13.66 | 1.47 | 67.83 | 22.13 | 18.16 | 7.07 | 1.09 | 14.12 | 5.02 | 3.02 | 7.86 | **9.59** | 15.87 |
| 2.13 | 28.72 | 70.77 | 13.56 | 1.47 | 67.11 | 21.73 | 17.82 | 6.81 | 2.22 | 18.60 | 8.85 | **1.08** | 5.61 | 12.86 | 19.67 |

**Summary**

In summary, we have done a numerical analysis of the BRs and direct CP-asymmetries of B decays to two light pseudoscalar nonets with $|\Delta S|=1$ using QCD factorization. We find that by properly choosing the phenomenological parameters $\rho_{H,A}$ and $\phi_{H,A}$, which parameterize logarithmic divergences in hard spectator scattering contribution and annihilation contribution, and also with minor adjustments of semileptonic form factors, $m_s$ and CKM phase angle $\gamma$, a good fit of majority of BRs as well as the largest direct CP-asymmetry observed in $B^- \to K^-\eta$ decay can be obtained. The small remaining discrepancies, we believe, may be removed by possible power corrections to penguin amplitudes including long distance charming penguins, final state interactions, penguin annihilation characterized by the parameters $\beta_3^{u,c}$[9] and higher order radiative corrections. We therefore conclude that at present there is no strong evidence for beyond the standard model physics.

**Acknowledgement:** JPS thanks authorities at University of Oregon, Eugene for hospitality, where part of the work was done. JPS also thanks Prof. S. K. Singh for fruitful discussion and Balkrishna Shah for help in the computational work. NGD thanks U. S. Department of Energy for support under contract no. DE-FG02-96ER40969.

**References**:

1. B. H. Behrens et al.[ CLEO Collaboration], Phys. Rev. Lett. **80**, 3710 (1998).
2. Y. Amhis et al. (Heavy Flavor Averaging Group), arXiv: 1207.1158.
3. M. Bauer and B. Stech, Phys. Lett. **B152**, 380 (1985); M. Bauer, B. Stech and M. Wirbel, Z. Phys. **C34**, 103 (1987).
4. M. Beneke, G. Buchala, M. Neubert and C. T. Sachrajda, Phys. Rev. Lett. **83**, 1914 (1999).
5. M. Beneke, G. Buchala, M. Neubert and C. T. Sachrajda, Nucl. Phys. **B606**,245 (2001).
6. Mao-Zhi Yang and Ya-Dong Yang, Nucl. Phys. **B609**, 469 (2001).
7. D. Du, H. Gong, J Sun, D. Yang and G. Zhu, Phys. Rev. **D65**, 074001(2002).
8. M. Beneke and M. Neubert, Nucl. Phys. **B651**, 225 (2003).
9. M. Beneke and M. Neubert, Nucl. Phys. **B675**, 333 (2003).
10. D. Du, J. Sun and G. Zhu, Phys. Rev. **D67**, 014023 (2003) [ arXiv: hep-ph/0209233].
11. B. Dutta, C.S. Kim, S. Oh and G. Zhu, Eur. Phys. J. **C37**, 273 (2004).
12. T. N. Pham, Phys. Rev. **D77**, 014024 (2008).
13. C.-H. V. Chang and H.-n. Li, Phys. Rev. **D55**, 5577 (1997); T.-W.Yeh and H.-n. Li, Phys. Rev. **D56**, 1615 (1997).
14. E. Kou and A.I. Sanda, Phys. Lett. **B525**, 240 (2002).
15. J.-F. Hsu, Y.-Y.Charng and H. Li, arXiv: 0711.4987.
16. Y.-Y. Fan, W.-F.Wang, S. Cheng and Z.-J.Xiao, Phys. Rev. **D87**, 094003 (2013).

17. T. Feldmann, P. Kroll and B. Stech, Phys. Rev. **D58**, 114006 (1998); T. Feldmann, P. Kroll and B. Stech, Phys. Lett. **B449**, 339 (1999).
18. A.G. Akeroyd, C.-H.Chen and C.-Q.Geng, Phys. Rev. **D75**, 054003 (2007).
19. J.-M. Gerard and E. Kou, Phys. Rev. Lett. **97**, 261804 (2006).
20. A. R. Williamson and J. Zupan, Phys. Rev. **D74**, 014003 (2006) [arXiv :hep-ph/0601214].
21. H. Y. Cheng, C. K. Chua and  A. Soni, Phys. Rev. **D72**, 014006 (2005).
22. Y-D. Tsai, H. Li and Q. Zhao, Phys. Rev. **D85**,034002 (2012).
23. G. Buchala, A. J. Buras and M. E. Lautenbacher, Rev. Mod. Phys. **68**, 1125 (1996).
24. J. Beringer et al. (Particle Data Group), Phys. Rev. **D86**, 010001 (2012).
25. K. Nakamura et al. [Particle Data Group Collaboration], J. Phys. **G37**, 075021 (2010).
26. J. Laiho, E. Lunghi and S. R. Van de Water, Phys. Rev. **D81**, 034503 (2010).
27. W. Luch, D. Melikhov and S. Simula, arXiv : 1305.7099.
28. Y.-Y. Charng, T. Kurimoto and H. Li, Phys. Rev. **D74**, 074024 (2006); hep-ph/0609165.
29. P.  Ball and R. Zwicky, Phys. Rev. **D71**, 014015 (2005).
30. H.-Y. Cheng and C.-K. Chua, arXiv: 1012.5504 (2010).
31. C.-K. Chua, Phys. Rev. **D78**, 076002 (2008).
32. J. P. Singh and J. Pasupathy, Phys. Rev.**D79**, 116005 (2009).
33. H.-Y. Cheng, H.-n.Li and K. -F. Liu, Phys. Rev. **D79**, 014024 (2009).